\documentclass{eas}
\usepackage{graphicx}
\begin{document}

\title{The Kinematics and Zero Point of the
log $P$ -- $<M_K>$ Relation for Galactic Field
RR Lyrae Variables via Statistical Parallax}
\author{A.~K.~Dambis}\address{Sternberg Astronomical Institute, Universitetskii pr. 13,
Moscow, 119992 Russia}
%
%
\begin{abstract}
The kinematical parameters of the local field RR Lyrae population
and the zero point of the log $P$--$<M_K>$ relation for these
variables are inferred by applying the statistical parallax
(maximum-likelihood) technique to a sample of 182 RR Lyraes with
known periods, radial-velocities, metallicities, K-band
photometry, and absolute proper motions on the ICRS system.
Hipparcos, Tycho-2, SPM, UCAC, NPM1, and the Four-Million Star
Catalog (Volchkov {\em et al. \/} \cite{volchkov}) were used as
the sources of proper motions; the proper motions of the last two
catalogs are reduced to the Hipparcos (ICRS) system (Dambis \&
Rastorguev \cite{dr}). The K-band magnitudes were adopted from
the list of Fernley {\em et al. \/} (\cite{fernley}) and
supplemented by the data of the 2MASS Second Incremental Data
Release. The parameters of the velocity distribution are found to
be ($U_0$ = -10 $\pm$ 10, $V_0$ = -51 $\pm$ 8, $W_0$ = -14 $\pm$
5) km/s, ($\sigma_U$ = 62 $\pm$ 10, $\sigma_V$ = 45 $\pm$ 8,
$\sigma_W$ = 28 $\pm$ 6) km/s
 and
($U_0$ = -23 $\pm$ 13, $V_0$ = -213 $\pm$ 12, $W_0$ = -5 $\pm$ 8)
km/s ($\sigma_U$ = 157 $\pm$ 12, $\sigma_V$ = 98 $\pm$ 8,
$\sigma_W$ = 91 $\pm$ 7) km/s for the thick-disk (41 stars) and
halo (141 stars) objects, respectively. The zero point of the
infrared PL relation of Jones {\em et al. \/} (\cite{jones})
(based on the results obtained using the Baade-Wesselink method)
is confirmed: we find $<M_K>$ = -2.33 $\cdot$ log $P_F$-0.82 $\pm
0.12$ compared to $<M_K>$ = -2.33 $\cdot$ log $P_F$-0.88 as
inferred by Jones {\em et al. \/} (\cite{jones}). A conversion of
the resulting log $P$-$<M_K>$ relation to $V$-band luminosities
yields the metallicity-luminosity relation $<M_V>$ = +1.04 + 0.14
$\cdot$ [Fe/H] $\pm$ 0.11.Our results imply a solar Galactocentric
distance of $R_0$ = 7.6 $\pm$ 0.4 kpc and an LMC distance modulus
of $DM_{LMC}$ = 18.18 $\pm$ 0.12 (cluster RR Lyraes) or
$DM_{LMC}$ = 18.11 $\pm$ 0.12 (field RR Lyraes), thereby favoring
the so-called short distance scale.
\end{abstract}
\maketitle
\section{Introduction}
RR Lyrae variables are among the tools of greatest value for
determining the distances to old stellar systems both in our
Galaxy and beyond. These stars are so important as one of the
distance indicators in the Universe, because they have very
similar luminosities (the dispersion of $V$-band absolute
magnitudes of RR Lyraes of the same metal abundance does not
exceed 0.15$^m$ (Sandage \cite{sandage})), are easily identifiable
(because of substantial variability), and reliably classifiable
(by low metallicity, period, and the shape of the light curve).

Since the pioneering work of Pavlovskaya (\cite{pavlovskaya}), who
was the first to apply the statistical-parallax technique to
estimate the mean absolute magnitude of RR Lyrae stars, this
method was, in increasingly sophisticated form, used by many
authors (Rigal \cite{rigal}; van Herk \cite{vh}; Hemenway
\cite{hemenway}; Clube \& Dawe \cite{cd}; Hawley {\em et al. \/}
\cite{hawley}; Strugnell {\em et al. \/} \cite{strugnell}; Layden
{\em et al.\/} \cite{layden96}; Popowski \& Gould \cite{pg98a},
\cite{pg98b}; Fernley {\em et al. \/} \cite{fernley}; Tsujimoto
{\em et al. \/} \cite{tsujimoto}; Dambis \& Rastorguev \cite{dr})
on ever increasing star samples.

However, all the above authors used statistical parallax
technique to determine the mean $V$-band absolute magnitude or to
set the zero point of the [Fe/H]-$<M_V>$ relation. Here we for
the first time use this method in its rigorous form (as described
by Murray (\cite{murray}) and Hawley {\em et al. \/}
(\cite{hawley})) to calibrate the infrared period-luminosity
relation (log $P_F$ - $<M_K>$) for RR Lyrae stars.

\section{Observational data}

I adopted the periods and pulsation modes from the General
Catalog of Variable Stars (Kholopov {\em et al. \/} \cite{gcvs}).
The fundamental-mode periods $P_F$ for first-overtone pulsators
(RRc type variables) were computed as log $P_F$ = log $P$ + 0.127
(see Frolov \& Samus' \cite{fs}). Our source of radial velocities,
[Fe/H], and interstellar extinction, $E_{B-V}$, was the work of
Beers {\em et al. \/} (2000). We collected $K$-band magnitudes
from two sources: the list of Fernley {\em et al. \/}
(\cite{fernley}) and (for the remaining stars), the 2MASS Second
Incremental Data Release. We adopted the absolute proper motions
of RR Lyrae variables from the Hipparcos (ESA \cite{hip}),
Tycho-2 (Hog {\em et al. \/} \cite{hog}), SPM (Platais {\em et
al. \/} \cite{platais}), UCAC (Zacharias et al.
\cite{zacharias}), NPM1 (Hanson {\em et al. \/} \cite{hanson}),
and Four-Million Star Catalog (Volchkov {\em et al. \/}
\cite{volchkov}). The proper motions of the latter two sources
are reduced to the Hipparcos (ICRS) frame and adopted from Dambis
\& Rastorguev (\cite{dr}).

\section{Disk/Halo Separation}

Layden (\cite{layden95}) and Layden {\em et al. \/}
(\cite{layden96}) showed that the kinematic population of RR
Lyraes in our Galaxy breaks conspicuously into two subclasses:
halo and thick-disk stars. At first approximation, the two
subpopulations can be separated by the metallicity value [Fe/H] =
-1.0: most of the more metal-deficient stars behave as halo
objects and most of more metal-rich stars, as thick-disk objects.
Here we subdivided the entire sample into halo and thick-disk
subsamples as described by Layden {\em et al. \/}
(\cite{layden96}) and Dambis \& Rastorguev (\cite{dr}).

\section{Initial distances}

Here we determined the initial distances to the RR Lyrae stars of
our sample based on the infrared PL relation by Jones {\em et al.
\/} (\cite{jones}) based on the application of Baade-Wesselink
method to field RR Lyrae stars:
\begin{equation}
 <M_{K(Jones)}> = -2.33 \cdot {\rm log} P_F - 0.82.
\end{equation}
We use the statistical-parallax technique to determine a
correction $\Delta M$ to the zero point of this relation so that
\begin{equation}
<M_{K(true)}> = <M_{K(Jones)}> + \Delta M.
\end{equation}

\section{Results}
\subsection{Period-Luminosity and Period-Metallicity Relations}

The results are summarized in Table~1. Here Halo-1,2,3/Disk-1,2,3
indicate the names of the halo/disk subsamples according to the
three partitions of the entire sample; $N$, the number of stars
in the subsample; ($U_0$,$V_0$, $W_0$) and ($\sigma_U$,
$\sigma_V$, $\sigma_W$) are the mean heliocentric velocity and
velocity dispersion components of the corresponding subsample,
and $\Delta M$, the absolute-magnitude correction mentioned above.

\begin{table}
  \centering
  \caption{Kinematical parameters and $K$-band absolute-magnitude correction of Galactic field RR Lyrae variables}\label{tab1}
\begin{tabular}{|r |r |r |r r r r r r |r|}
\hline
  Sample & $N$ & ${\rm <[Fe/H]>}$ & $U_0$ & $V_0$ & $W_0$ & $\sigma U$ & $\sigma_V$ & $\sigma W$ & $\Delta M$ \\
  \hline
    &   &   &   &   &   & km/s  &   &   &   \\ \hline
  Halo-1 & 148 & -1.57 & -23 & -207 & -5 & 154 & 99 & 89 & +0.07 \\
        &    &      & $\pm$ 13 & $\pm$ 12 & $\pm$ 7 & $\pm$ 12 & $\pm$ 7 & $\pm$ 6 & $\pm$ 0.12 \\
    &   &   &   &   &   &   &   &   &   \\
Halo-2 & 152 & -1.56 & -23 & -202 & -4 & 152 & 101 & 88 & +0.07 \\
        &    &      & $\pm$ 13 & $\pm$ 11 & $\pm$ 7 & $\pm$ 12 & $\pm$ 8 & $\pm$ 6 & $\pm$ 0.12 \\
    &   &   &   &   &   &   &   &   &   \\
Halo-3 & 141 & -1.58 & -23 & -213 & -5 & 157 & 98 & 91 & +0.06 \\
        &    &      & $\pm$ 13 & $\pm$ 12 & $\pm$ 8 & $\pm$ 12 & $\pm$ 8 & $\pm$ 7 & $\pm$ 0.12 \\
    &   &   &   &   &   &   &   &   &   \\

  Disk-1 & 34 & -0.58 & -6 & -38 & -12 & 61 & 38 & 26 & +0.26 \\
        &    &      & $\pm$ 11 & $\pm$ 7 & $\pm$ 5 & $\pm$ 10 & $\pm$ 7 & $\pm$ 6 & $\pm$ 0.31   \\
    &   &   &   &   &   &   &   &   &   \\
  Disk-2 & 30 & -0.51 & -4 & -41 & -16 & 63 & 43 & 23 & +0.09 \\
        &    &      & $\pm$ 12 & $\pm$ 9 & $\pm$ 5 & $\pm$ 12 & $\pm$ 9 & $\pm$ 6 & $\pm$ 0.32   \\
    &   &   &   &   &   &   &   &   &   \\
Disk-3 & 41 & -0.72 & -10 & -51 & -14 & 62 & 45 & 28 & +0.01 \\
        &    &      & $\pm$ 10 & $\pm$ 8 & $\pm$ 5 & $\pm$ 10 & $\pm$ 8 & $\pm$ 6 & $\pm$ 0.26   \\
    &   &   &   &   &   &   &   &   &   \\ \hline
\end{tabular}
\end{table}

\ The inferred kinematical parameters of RR Lyrae subpopulations
agree well with those determined by Dambis \& Rastorguev
(\cite{dr}) proceeding from the [Fe/H] - $<M_V>$ relation. The
resulting absolute-magnitude correction $\Delta M$ = +0.06 $\pm$
0.12 (based on the most homogeneous Halo-3 sample) implies a
final PL relation of:

$<M_K>$ = -2.33 $\cdot$ log $P_F$ - 0.82 $\pm$ 0.12.

This relation can be converted into a [Fe/H]-$<M_V>$ relation.
This is done by computing $<M_V>$ for stars with known $<M_K>$
(see above):
\begin{equation}
<M_V> = <M_K> + <V_0>-<K_0>,
\end{equation}
where
\begin{equation}
<V_0> - <K_0> = <V> - <K> -(A_V - A_K)
\end{equation}
and
\begin{equation}
A_V - A_K = (3.1 - 0.1 \cdot 3.1) \cdot E_{B-V} = 2.79 \cdot
E_{B-V},
\end{equation}
and fitting the resulting $<M_V>$ to a relation $<M_V>$ = $a$ +
$b$ $\cdot$ [Fe/H]. We thus obtain:
\begin{equation}
<M_V> = +1.04 + 0.14 \cdot [Fe/H] \pm 0.10
\end{equation}
(based on 107 stars with both $K$-band and photoelectric $V$-band
photometry).

\subsection{The Distance to the Galactic Center}

Carney {\em et al. \/} (\cite{carney}) obtained a solar
Galactocentric distance of $R_0$ = 7.8 $\pm$ 0.4 kpc based on IR
photometry of 58 RR Lyraes in the Galactic bulge and the PL
relation of Jones {\em et al. \/} (\cite{jones}). Our distance
scale is 0.06$^m$ fainter, implying a solar Galactocentric
distance of: $R_0$ = 7.6 $\pm$ 0.4 kpc based on the
statistical-parallax zero point of the IR PL relation for RR
Lyrae variables. This result favors the so-called short distance
scale.

\subsection{The Distance Modulus of the LMC}

Our [Fe/H] - $<M_V>$ relation (converted from log $P_F$ - $<M_K>$
relation) yields a mean distance modulus of $DM_{LMC(cluster)}$ =
18.18 $\pm$ 0.11 for RR Lyraes of seven LMC globular clusters
(based on the data adopted from Walker (\cite{walker})) and
$DM_{LMC(field)}$ = 18.11 $\pm$ 0.11 for field RR Lyrae (based on
OGLE survey  data from Udalski (\cite{udalski})). The inferred
LMC distance modulus again corroborates the short distance scale.

\section{Acknowledgments}
This publication makes use of data products from the Two Micron
All Sky Survey, which is a joint project of the University of
Massachusetts and the Infrared Processing and Analysis
Center/California Institute of Technology, funded by the National
Aeronautics and Space Administration and the National Science
Foundation.

The work was supported by the Russian Foundation for Basic
Research, grants nos. 01- 02-06012, 00-02-17804, 99-02-17842, and
01-02-16086; Astronomy State Research and Technology Program; and
the Council for the Support of Leading Scientific Schools, grant
no. 00-15-96627.


\end{document}